\documentclass[%
 reprint,
 amsmath,amssymb,
 aps,
 prl,
]{revtex4-2}

\usepackage{graphicx}
\usepackage{dcolumn}
\usepackage{bm}

\usepackage{xcolor} 

\begin{document}

\preprint{APS/123-QED}


\title{Turbulence-Generated Stepped Safety Factor Profiles in Tokamaks with Low Magnetic Shear}

\author{Arnas Vol\v{c}okas}
 \email{Arnas.Volcokas@epfl.ch}
\author{Justin Ball}%
\author{Giovanni Di Giannatale}%
\author{Stephan Brunner}%
\affiliation{%
 Ecole Polytechnique F\'{e}d\'{e}rale de Lausanne (EPFL), Swiss Plasma Center (SPC), CH-1015 Lausanne, Switzerland 
}%

\date{\today}

\begin{abstract}
Nonlinear local and global gyrokinetic simulations of tokamak plasmas demonstrate that turbulence-generated currents flatten the safety factor profile near low-order rational surfaces when magnetic shear is low, even when the plasma $\beta$ is small.
A large set of flux tube simulations with different safety factor profiles (e.g. linear and non-linear safety factor profiles) and global simulations with reversed magnetic shear profiles show that such stepped safety factor profiles dramatically reduce the heat transport and are a robust phenomenon. 
This mechanism may play a key role in the triggering of internal transport barriers (ITBs) and more generally reveal novel strategies for improving confinement in devices with low magnetic shear.

\end{abstract}

\maketitle

\paragraph*{Introduction. ---}

Reducing energy losses, thereby increasing the confinement time, is critical for the design of economically viable magnetic confinement fusion power plants \cite{Freidberg2007, Fasoli2023_essay}. 
An internal transport barrier (ITB) represents an advanced confinement scenario that offers a reduced heat diffusivity in the plasma core, a large bootstrap current fraction and favorable power exhaust \cite{Wolf2003, Connor_2004_ITB, Ida2018}. 
ITBs are most readily identified by a localized steepening of the pressure profile and reduction in turbulent transport levels in the plasma core. 
Although ITBs have been experimentally observed in almost all major tokamaks, including JT-60U \cite{Koide1994_ITBatJT60U}, TFTR \cite{Levinton1995_ITBatTFTR}, DIII-D \cite{Strait1995_ITBatDIIID}, JET \cite{Crisanti2002_ITBatJET}, ASDEX-U \cite{Hobrik2001_ITBatASDEXU}, EAST \cite{Li2022_ITBatEAST}, KSTAR \cite{Chung2017_ITBonKSTAR}, and TCV \cite{Coda2019_ITBonTCV}, the fundamental mechanisms underlying their formation remain poorly understood.

In this Letter, we present a numerical study of plasma turbulence self-organization under conditions relevant to ITB formation, specifically near surfaces with rational values of the safety factor $q$ and low magnetic shear $\hat{s}$ \cite{Eriksson2002, Shafer2009_q2_ITB, Koide_1994_q3_ITB_JT60}.
Here, $q$ is the number of toroidal turns a field line makes around the torus per poloidal turn and $\hat{s}=(r/q)(dq/dr)$ is its associated shear, where $r$ is the radial variable.
A magnetic surface is called ``rational" if the safety factor can be expressed as an irreducible fraction $q=m/n$ (where $n, m \in \mathbb{Z}^{+}$) of ``order" $n$.
We demonstrate a novel turbulence self-organization mechanism that occurs through the modification of the safety factor profile.
Due to strong parallel self-interaction (i.e. turbulent eddies ``biting their own tail" in the parallel direction \cite{Ball2020}) at low magnetic shear $\hat{s} \ll 1$, 
radial inhomogeneity in turbulence drives electric currents through the redistribution of electron momentum. 
This flattens the imposed safety factor profile around rational surfaces, creating extended regions of near-zero magnetic shear around these surfaces, which further strengthen self-interaction.
The outcome of this process is a significant turbulence transport reduction. 
The positive feedback loop between the safety factor profile modifications and parallel turbulence self-interaction can explain several experimental observations, including ITB splitting in JET \cite{Joffrin2002} and early barrier formation in DIII-D \cite{Austin2006_ITBnearintegerq}.


In this study, we employ gyrokinetics, a kinetic model of magnetized plasma turbulence \cite{Catto1978, Frieman1982, Brizard2007, Abel2013}, and solve the coupled system of gyrokinetic Vlasov-Maxwell equations numerically using the  continuum flux tube code GENE \cite{Jenko2000, Gorler2011} and the particle-in-cell (PIC) global code ORB5 \cite{Lanti_2020_ORB5}.
The main numerical results presented in this Letter were obtained using collisionless, weakly electromagnetic (EM) flux tube simulations with a full kinetic treatment of both ions and electrons. 
In flux tube simulations \cite{Beer1995}, we use a computational domain in field aligned coordinates $\left( x,y,z \right) \in \left[ 0, L_{x} \right] \times \left[ 0, L_{y} \right] \times \left[ - \pi N_{pol}, \pi N_{pol} \right]$ that is narrow in the perpendicular directions but extended along the magnetic field lines. 
Here $L_{x}$ and $L_{y}$ are the radial ($x$) and binormal ($y$) domain widths respectively, $z$ is the parallel coordinate and $N_{pol}$ quantifies the parallel length by the number of poloidal turns around the torus.
We focus on the Ion Temperature Gradient (ITG)-driven turbulence regime.
In some cases the electron heat flux matches or exceeds the ion heat flux, suggesting a mixed Trapped Electron Mode (TEM)-ITG regime.

\paragraph*{Turbulent Modifications of Safety Factor Profile ---}

Previous work on turbulence near rational surfaces has shown that kinetic electron physics leads to radial corrugations of plasma profiles both in local \cite{Dominski2015, Ball2020, Ajay2020} and global \cite{Waltz2006,Dominski_2017_corrugations} gyrokinetic simulations.
These corrugations result from parallel turbulent self-interaction \cite{Ball2020}.

Specifically, parallel self-interaction induces radial variation in turbulence characteristics, which generates the local corrugations in the plasma profiles. 
This mechanism is analogous to the momentum flux driven by the variation in turbulence characteristics discussed in Refs. \cite{Parra2015a, McDevitt2017}.
Importantly, self-interaction becomes even stronger at low magnetic shear as turbulent eddies extend further along magnetic field lines \cite{Volcokas2023, Volcokas2024_Part1_L, Volcokas2024_Part2_NL}.

In the present study, we show for the first time that such corrugations in the parallel current $j_{\parallel}$ enable a feedback loop that dramatically enhances self-interaction.
Parallel currents generate a steady zonal parallel magnetic potential perturbation $A_{\parallel}$ through Amp\`{e}re's law, which corresponds to a safety factor profile modulation of
\begin{equation}
\label{eq:q_Apar_final}  
    \Tilde{q}_{A_{\parallel}}(x) \propto \partial_{x} \langle \langle A_{\parallel} \rangle_{FS} \rangle_{t}.
\end{equation}
Here $\langle ... \rangle_{FS}$ is a flux surface average and $\langle ... \rangle_{t}$ is a time average over multiple eddy correlation times.

Additionally, to study the impact of turbulence-generated safety factor modulations, it will be useful in some cases to externally impose such modulations.
This is done in flux tube simulations using a periodic safety factor modulation $\tilde{q}(x)$, defined in terms of magnetic shear Fourier coefficients $\tilde{s}_{n}^{S}$ and $\tilde{s}_{n}^{C}$  \cite{Ball2022}.

Once we take into account the modulations, the total safety factor profile can be written as 
\begin{equation}
\label{eq:total_mod_q}
    q_{tot}(x) = q(x) + \Tilde{q}_{A_{\parallel}}(x) = q_{0}+ \frac{q_{0}}{r_{0}} \hat{s} x + \Tilde{q}(x)+\Tilde{q}_{A_{\parallel}}(x),
\end{equation}
which combines the imposed linear $q_{0}+(q_{0}/r_{0})\hat{s}x$ and non-uniform $\Tilde{q}(x)$ safety factor profile terms with a turbulent-generated contribution $\Tilde{q}_{A_{\parallel}}(x)$.
Here $q_{0}$ is the safety factor value at the center $r_{0}$ of the radial domain.

\paragraph*{Linear shear. ---}

We begin by investigating how standard linear safety factor profiles are modified by turbulence-generated currents.
We examine scenarios with low, but finite, background magnetic shear $\hat{s}$, focusing on $|\hat{s}| = 0.1$ with no external modulations $\Tilde{q}(x) = 0$.
In our simulations, we consider two types of turbulence drives: CBC-like \cite{Dimits2000}, and pure ITG (pITG)\cite{Volcokas2024_Part1_L}, which involves only a finite ion temperature gradient.
Our focus is primarily on CBC, as it represents a less idealized scenario.

\begin{figure}[hbt]
\includegraphics[width=0.5\textwidth]{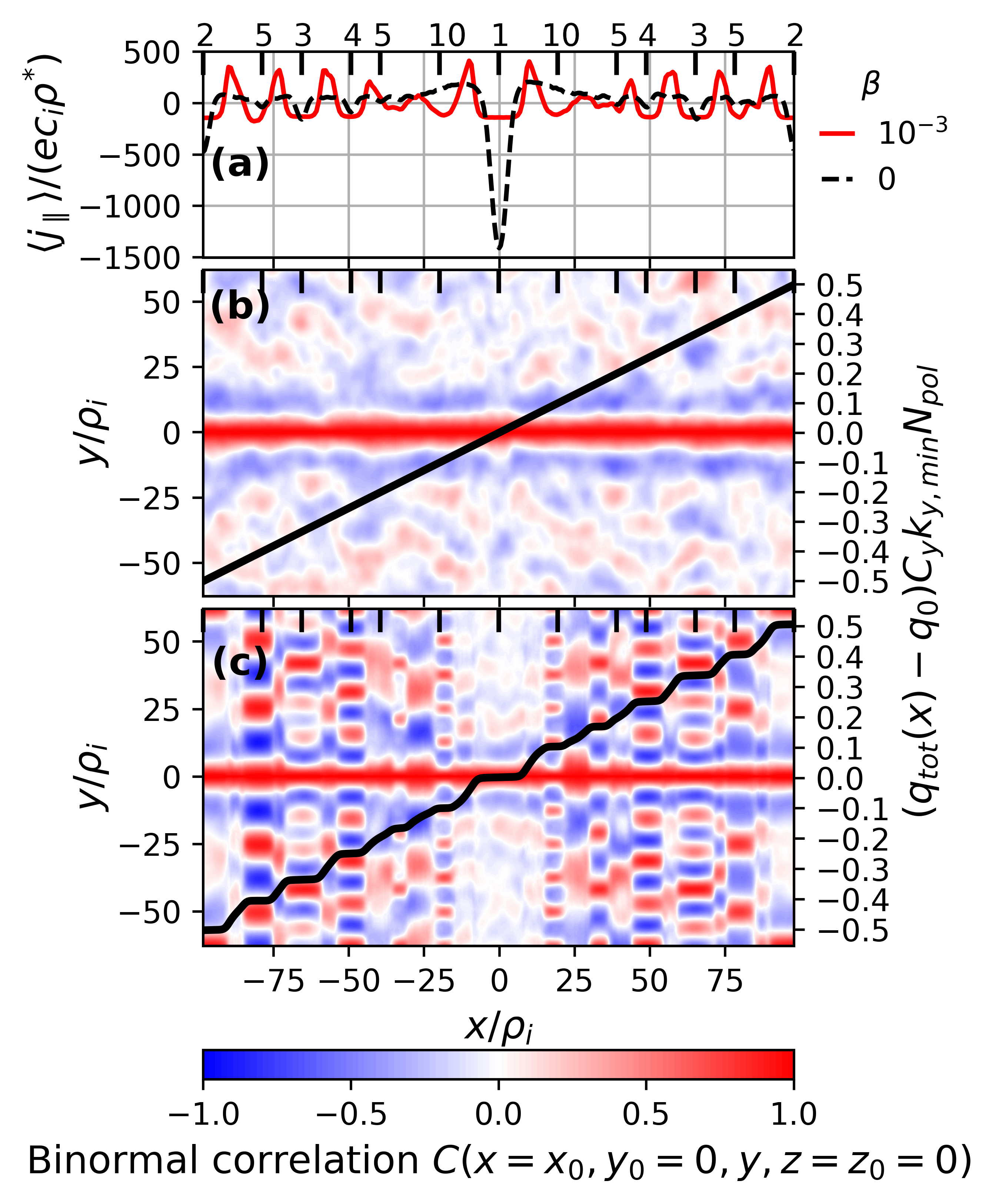}
\caption{\label{fig:PRL_QFlat_s01_biCorr_b0vsb0001} (a) Radial profile of the flux surface-averaged parallel current for ES (black) and weakly EM (red) simulations, with a background magnetic shear $\hat{s}=0.1$. The lower subplots display the binormal ES potential auto-correlation (color plot) as a function of radial position as well as the safety factor profile (black) for (b) $\beta=0$ and (c) $\beta=10^{-3}$. The top axes show the order and positions of selected rational surfaces. Here $\rho_{i}$ is the ion gyroradius, $c_{i}$ is the speed of sound and $C_{y} \approx r_{0}/q_{0}$. In these simulations $N_{pol} = 1$.}
\end{figure}

Near the rational surfaces of the flux tube \cite{Ball2020}, we observe stationary zonal parallel current profile corrugations due to turbulent current drive, as shown in Fig. \ref{fig:PRL_QFlat_s01_biCorr_b0vsb0001}(a).
We find that when magnetic shear is small, the resulting steady zonal magnetic potential can significantly alter the background linear safety factor profile, leading to a stepped profile with flattening at multiple low-order rational surfaces.
Importantly, even purely electrostatic (ES) turbulence drives current, though EM effects are necessary for the turbulent current to modify the background $q$ profile. 
When the plasma $\beta \neq 0$ (plasma pressure over magnetic pressure), the currents lead to stationary magnetic potential corrugations, causing safety factor profile to change via Eqs. \eqref{eq:q_Apar_final} and \eqref{eq:total_mod_q}.


Fig. \ref{fig:PRL_QFlat_s01_biCorr_b0vsb0001}(b, c) shows the total safety factor profiles for ES $\beta = 0$ and EM $\beta=10^{-3}$ cases, where we see a stepped safety factor profile with flattening at rational surfaces for the EM $\beta=10^{-3}$ case.
Note that the right vertical axis is normalized such that a value of $(q_{tot}(x)-q_{0})C_{y}k_{y, min}N_{pol} =1/n$ indicates an $n^{th}$ order rational value of the safety factor and $0$ corresponds to an integer surface.
Here $k_{y,min} = 2\pi/L_{y}$ is the minimal binormal wave number.
We find that the direction of the currents flips for simulations with $\hat{s} = - 0.1$, so that flattening still occurs at rational surfaces.

At radial locations where the safety factor profile is flattened, the characteristics of turbulent eddies undergo substantial changes.
When the total magnetic shear is zero, turbulent eddies can extend along the magnetic field lines for hundreds of poloidal turns, enhancing turbulent self-interaction \cite{Volcokas2023}.
At certain radial locations one consequence of this stronger self-interaction is ``eddy squeezing", which is known to lead to reduced turbulent transport \cite{Volcokas2023, Volcokas2024_Part2_NL}.
This effect can be visualized using the two-point Eulerian correlation $C(x_{0}, y_{0}, z_{0}, x, y, z)$ \cite{Wallace2014, Ball2020} at the outboard midplane $z=z_{0}=0$ as a function of radius $x=x_{0}$.
For the $\beta=10^{-3}$ case, we can easily identify $1^{st}$, $2^{nd}$, $3^{rd}$, $4^{th}$, $5^{th}$ and even $10^{th}$ order rational surfaces by the number of peaks in the binormal direction.
This corresponds to the number of times individual turbulent eddies pass through the parallel boundary, each time being shifted in the birnormal direction, before ``biting their own tail".
The corrugated pattern in the correlation and the stepped safety factor profile show that turbulence-generated currents have a profound feedback effect on turbulence when magnetic shear is low.

\begin{figure}[hbt]
\includegraphics{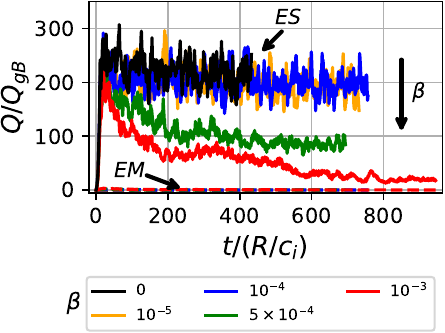}
\caption{\label{fig:PRL_s01_Qtrace_betaScan_050924} Time traces of ES (solid) and EM (dashed) heat fluxes for CBC with fixed $\hat{s}=0.1$ and varying values of $\beta$.}
\end{figure}

\begin{figure}[hbt]
\includegraphics[width=0.5\textwidth]{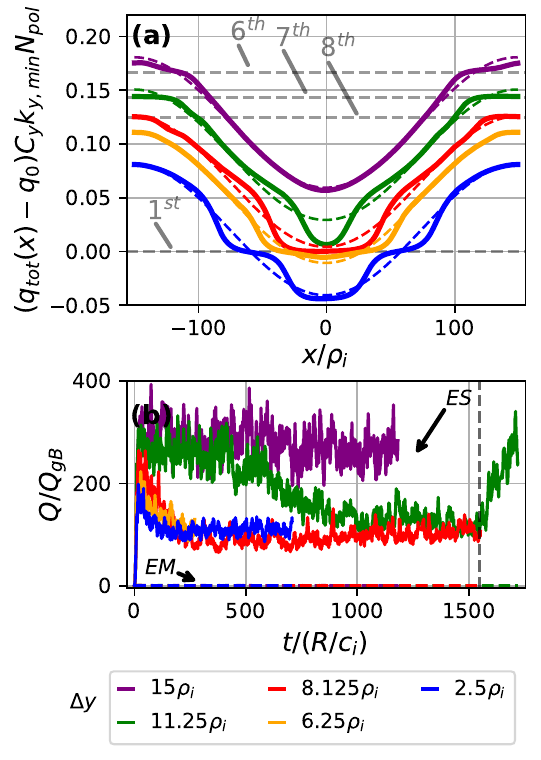}
\caption{\label{fig:PRL_s0_sSm0025_qprofile_Qtrace} (a) A safety factor profile scan varying the binormal shift $\Delta y$ for a weakly EM $\beta=10^{-4}$ case with a sinusoidal safety factor profile. The solid curves are the turbulence-modified profiles, while the dashed curves indicate the initial profiles. The grey dashed horizontal lines denote the order of various rational surfaces (as indicated in the plot). (b) Heat flux time traces with varying binormal shift $\Delta y$.
After the grey broken line, the $\Delta y = 11.25 \rho_{i}$ (green) simulation imposes $\langle A_{\parallel} \rangle_{y} = 0$. The ES heat flux (solid) always dominates over the EM contribution (dashed).}
\end{figure}

This modification of the safety factor profile by the steady zonal $A_{\parallel}$ leads to turbulence stabilization.
As shown in Fig. \ref{fig:PRL_s01_Qtrace_betaScan_050924}, the total heat flux is significantly reduced at higher $\beta$.
We attribute this primarily to the steady zonal component of $A_{\parallel}$.
Artificially setting the zonal $A_{\parallel}$ to zero causes heat fluxes to increase and match the ES case (as will be shown in Fig. \ref{fig:PRL_s0_sSm0025_qprofile_Qtrace}(b)).
We also modified the radial safety factor profile in an ES ($\beta = 0$) simulation by including an appropriate $\Tilde{q}(x)$ to match the final steady $q_{tot}(x)$ profile from the self-consistent EM ($\beta = 10^{-3}$) simulation and observed reduced heat fluxes, quantitatively matching with the EM result.
These findings confirm that the steady zonal $A_{\parallel}$ (and the corresponding safety factor profile modification) is the main stabilizing factor.

To further understand the stabilization mechanism, we conducted adiabatic electron simulations comparing a standard linear $q$ profile with an imposed corrugated profile taken from kinetic electron EM simulations.
Adiabatic electrons eliminate strong self-interaction by reducing the parallel eddy length, allowing us to isolate the effects of safety factor profile curvature.
Using $\Tilde{q}(x)$, we imposed a $q_{tot}(x)$ profile similar to that from Fig. \ref{fig:PRL_QFlat_s01_biCorr_b0vsb0001} (c) and found it did not change the heat flux, suggesting that the modulation of the safety factor profile has minimal direct impact.
Similarly, linear kinetic electron simulations show that imposing a corrugated profile $\Tilde{q}(x)$ does not significantly change the linear mode growth rates. 
Therefore, we conclude that the substantial reduction in transport with $\beta$ in Fig. \ref{fig:PRL_s01_Qtrace_betaScan_050924} is due to the broad zero magnetic shear regions at rational surfaces enhancing turbulent eddy self-interaction.

\paragraph*{Non-linear safety factor profiles. ---}

To complement the previous section, we carried out a series of simulations with no background magnetic shear $\hat{s} = 0$ and an imposed sinusoidal safety factor profile modulation $\tilde{q}(x) \propto \hat{s}^{S}_{1} \cos{(2\pi x/L_{x})}$ of varying amplitude.

We found that, if the non-uniform shear amplitude $\Tilde{s}^{S}_{1}$ is small enough to include only the lowest-order rational $q$ value in the simulation domain, turbulent-generated currents completely flattened the imposed safety factor profile modulation. 
However, when the non-uniformity is substantial enough to introduce additional low-order rational surfaces, the safety factor profile was pulled towards and flattened around multiple rational $q$ values, as shown in Fig. \ref{fig:PRL_s0_sSm0025_qprofile_Qtrace} (a).
This results in a stepped safety factor profile similar to our standard simulations shown in Fig. \ref{fig:PRL_QFlat_s01_biCorr_b0vsb0001} (c). 
Furthermore, by introducing a binormal shift $\Delta y$ into the parallel boundary condition of the flux tube, leading to an additional offset $\Delta q \propto \Delta y$ \cite{Volcokas2024_Part1_L} of the imposed safety factor profile $q(x)$, we can perform a scan in $q(x)$ across an integer surface.
This approach allows us to emulate experimental conditions associated with ITB formation and search for changes in turbulence characteristics as the minimum safety factor crosses a rational $q$ value.

We performed a $\Delta y$ scan for a case with $\hat{s}^{S}_{1}=-0.025$, no background magnetic shear $\hat{s} = 0$, $L_{x}=300 \rho_{i}$, $\beta=10^{-4}$ and heavy electrons $m_{i}/m_{e} = 364$. 
We began the scan with the imposed safety factor profile significantly above an integer surface, lowering it to include the integer surface, and further decreasing it to include the integer value at two widely separated radial locations. 
The initial and turbulence-modified $q$ profiles from this scan are shown in Fig. \ref{fig:PRL_s0_sSm0025_qprofile_Qtrace}(a).

Several features of the modified safety factor profile are noteworthy. 
First, the profile flattens around rational surfaces, evident at the $1^{st}$ (integer) as well as the 6$^{th}$, 7$^{th}$ and 8$^{th}$ order surfaces. 
Second, even when the initial profile does not include a rational value, the modified profile is ``pulled" towards the nearest rational surface, as seen in the cases with $\Delta y = 8.125 \rho_{i}$ and $\Delta y = 11.25 \rho_{i}$. 
This indicates that, as the safety factor profile minimum approaches a low-order rational value, turbulent currents can locally modify the safety factor profile towards this value, potentially aiding in the formation of an ITB.
This can explain how DIII-D observed the effect of rational surfaces on ITB triggering before the rational surface was expected to have entered the plasma \cite{Austin2006_ITBnearintegerq}.

The heat flux time traces for these simulations are shown in Fig. \ref{fig:PRL_s0_sSm0025_qprofile_Qtrace}(b). 
The three simulations with initial profiles close to or crossing the lowest order rational surface have about three times lower total heat flux than the $\Delta y =15 \rho_{i}$ simulation, which is not close to any low-order rational surface.
Crucially, the $\Delta y = 11.25 \rho_{i}$ simulation initially exhibits high heat flux, which decreases as turbulence-generated currents build up and the safety factor profile evolves toward the lowest order rational surface.
If $\langle A_{\parallel} \rangle_{y}$ is set to zero in this simulation, the turbulent modification to the safety factor profile is eliminated $\Tilde{q}_{A_{\parallel}} = 0$, and, as expected, we see that the heat flux increases.

Overall, these simulations demonstrate a strong reduction in transport as a reversed shear safety factor profile crosses a low-order rational surface, consistent with experimental observations of ITBs. 
They further demonstrate that, while the main stabilizing effect arises from the safety factor profile's minimum coinciding with an integer value, this stabilization can be facilitated by turbulence-generated changes to the safety factor profile.

\paragraph*{Global simulations. ---}

To further demonstrate the experimental applicability of our flux tube simulations, we also performed global ORB5 simulations.
These simulations employed a reversed shear safety factor profile with the minimum $q_{\text{min}}$ located at $s = 0.51$.
We consider kinetic electrons with the mass ratio $m_{i}/m_{e} = 500$.
These global simulations model the full radial domain from $s=0$ to $s=1$, where the considered radial variable is $s = \sqrt{\Psi_{P} / \Psi_\text{P,edge}}$, $\Psi_\text{P}$ is the poloidal flux and $\Psi_\text{P,edge}$ is the value at the edge.
We studied DIII-D ($\rho_{*} \sim 1/180$) \cite{DIIID2021_DIIID} and TCV ($\rho_{*} \sim 1/100$) \cite{Hofmann1994_TCV} scale machines using both gradient-driven and flux-driven simulations, with $\rho_{*}$ defined at the $q_{min}$ location.
Here $\rho_{*} = \rho_{i}/a$, where $a$ is the device minor radius.
ES and weakly EM simulations were compared.
Additional simulation details will be made available in a future publication.

\begin{figure}[hbt]
\includegraphics{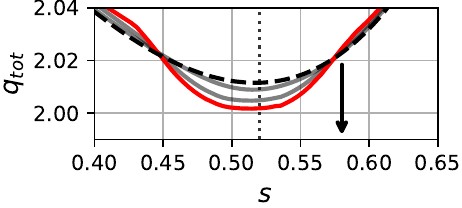}
\caption{\label{fig:GDiG_qEvolution_ZonalFLows_TCV_q2vsq201_FD_PRL_160125} Evolution of the safety factor profile from an initial profile with $q_{min}=2.01$ (dashed black) to a final profile with $q_{min} \approx 2$ (solid red) in EM TCV-scale flux-driven global ORB5 simulations.}
\end{figure}

We found excellent qualitative agreement with flux tube GENE results.
A quantitative comparison was not attempted due to the inherent differences in the models.
DIII-D-scale simulations with $q_{\text{min}} = 2$ exhibited much lower transport compared to those with $q_{\text{min}} = 2.03$ or $2.5$, aligning with the expected stabilizing effect of integer $q_{\text{min}}$.
Most importantly, a TCV-scale EM simulation with $q_{\text{min}} = 2.01$ also showed a significant reduction in transport.
This reduction is attributed to strong self-interaction near the integer surface further enhanced by turbulent current-induced modifications of the safety factor profile.
Fig. \ref{fig:GDiG_qEvolution_ZonalFLows_TCV_q2vsq201_FD_PRL_160125} shows how the minimum initially at $q_{min} = 2.01$ is ``pulled" towards $q_{\text{min}} = 2$ by turbulence, leading to the observed improvement in confinement.
These changes in the safety factor profile were analogous to the green case in Fig. \ref{fig:PRL_s0_sSm0025_qprofile_Qtrace}.

\paragraph*{Conclusions. ---}


Based on the results presented in this Letter, we propose that turbulence-generated currents are key mechanism in the triggering and formation of ITBs. 
They create modifications to the safety factor profile even when $\beta$ is low, leading to extended radial zones of near-zero magnetic shear around low-order rational surfaces. 
These zones allow turbulent eddies to extend further along the field line, resulting in stronger parallel self-interaction that reinforces the process. 
This leads to changes in plasma profiles and stabilizes turbulence.
While this study focused on tokamaks, the results are also expected to be important for modeling turbulence in stellarators, where magnetic shear can be low across large radial regions of the device, such as in HSX \cite{Gerard_2023_HSX} and W7-X \cite{Klinger_2019_W7X}.


\begin{acknowledgments}
The authors would like to thank Laurent Villard, Oleg Krutkin, Alessandro Geraldini, Ben McMillan, and Antoine Hoffmann for useful discussions pertaining to this work.
This work has been carried out within the framework of the EUROfusion Consortium, via the Euratom Research and Training Programme (Grant Agreement No 101052200 — EUROfusion) and funded by the Swiss State Secretariat for Education, Research and Innovation (SERI). Views and opinions expressed are however those of the author(s) only and do not necessarily reflect those of the European Union, the European Commission, or SERI. Neither the European Union nor the European Commission nor SERI can be held responsible for them.
We acknowledge the CINECA award under the ISCRA initiative, for the availability of high performance computing resources and support. 
This work was supported by a grant from the Swiss National Supercomputing Centre (CSCS) under project ID s1050, s1097 and s1252.

\paragraph*{Data availability statement. ---}
The data cannot be made publicly available upon publication
because of lab policy. The data that support the findings of this
study are available upon reasonable request from the authors.

\end{acknowledgments}

\bibliography{prlref}

\end{document}